\documentclass[nofootinbib,twocolumn, showpacs, preprintnumbers, amsmath, amssymb]{revtex4}

\usepackage{graphicx}
\usepackage{dcolumn}
\usepackage{bm}
\usepackage{mathbbol}
\usepackage{epsf}
\usepackage{mathrsfs}
\usepackage{slashed}
\usepackage{stmaryrd}
\usepackage{color}

\newcommand{\be}{\begin{equation}}
\newcommand{\ee}{\end{equation}}
\newcommand{\bea}{\begin{eqnarray}}
\newcommand{\eea}{\end{eqnarray}}
\newcommand{\bml}{\begin{mathletters} \baselineskip 10pt}
\newcommand{\eml}{\baselineskip 12pt \end{mathletters}}

\newcommand{\bra}{\langle}
\newcommand{\ket}{\rangle}

\def\lambdabar{\protect\@lambdabar}
\def\@lambdabar{%
\relax \bgroup
\def\@tempa{\hbox{\raise.73\ht0
\hbox to0pt{\kern.2\wd0\vrule width.7\wd0
height.1pt depth.1pt\hss}\box0}}%
\mathchoice{\setbox0\hbox{$\displaystyle\lambda$}\@tempa}%
{\setbox0\hbox{$\textstyle\lambda$}\@tempa}%
{\setbox0\hbox{$\scriptstyle\lambda$}\@tempa}%
{\setbox0\hbox{$\scriptscriptstyle\lambda$}\@tempa}%
\egroup }

\newcommand{\sfrac}[2]{{\textstyle \frac{#1}{#2}}}

\newcommand{\vc}[1]{\mbox{\boldmath$#1$}}

\newcommand{\dprod}[2]{#1 \! \cdot \! #2}

\begin{document}

\title{A Lorentz and gauge invariant measure of laser intensity}

\author{Thomas Heinzl}\email{theinzl@plymouth.ac.uk}

\author{Anton Ilderton}\email{abilderton@plymouth.ac.uk}

\affiliation{School of Mathematics and Statistics, University of
Plymouth\\
Drake Circus, Plymouth PL4 8AA, UK}

\date{\today}

\begin{abstract}
Focussing on null fields as simple models of laser beams we discuss the classical relativistic motion of charges in
strong electromagnetic fields.  We suggest a universal, Lorentz
and gauge invariant measure of laser intensity and explicitly
calculate and interpret it for crossed
field, plane wave and vortex models.
\end{abstract}

\pacs{42.55.Ah}



\maketitle

\section{\label{sec:1}Introduction}
Since the advent of chirped pulse amplification \cite{strickland:1985} optical lasers have reached
unprecedented intensities, the present state of the art being
about 10$^{22}$ W/cm$^2$. The associated electric fields of
approximately 10$^{13}$ V/m have renewed interest in the idea
of particle acceleration using laser fields. Using plasma wake
fields impressive progress has been made during the last few
years \cite{Geddes:2004tb}.  In vacuum, however, achieving laser acceleration is nontrivial as a laser field is oscillatory; the periodic sign change
seems to render acceleration of charges
unfeasible. Indeed there is a result named the Lawson-Woodward
theorem \cite{woodward:1946} which states that a
plane wave cannot accelerate charges (see \cite{palmer:1987}
for a thorough discussion). This is somewhat academic
to the extent that a laser beam is only crudely modelled
by a plane wave, as there is no ponderomotive force. Nevertheless, adopting plane waves as a first
approximation, it suggests that a laser may not be a very
efficient particle accelerator in vacuum -- unless its oscillatory and/or
plane wave characteristics can be overcome or modified. Part of
this paper will be to analytically assess some options
in this direction.

The most important parameter characterising the `strength' of a
high power laser is the `dimensionless laser amplitude' which,
in a particle physics context, may be written
\cite{bamber:1999}
\be \label{A0DEF}
  a_0 \equiv  \frac{e E_{\mathrm{rms}}}{\omega\, m_e c}\;,
\ee
with $e$ being the elementary charge, $E_{\mathrm{rms}} \equiv
\bra E^2\ket^{1/2}$ the rms electric field, $\omega$ the laser
frequency and $m_e$ the electron mass. The analogue of $1/a_0$ in atomic physics is the Keldysh parameter, see \cite{Dunne:2004nc} and references therein. Employing the laser wavelength $\lambdabar_L \equiv c/\omega$ we may write
\be\label{A0RATIO}
	 a_0=\frac{eE_{\mathrm{rms}} \lambdabar_L}{m_e c^2} \;,
\ee
which describes $a_0$ as the ratio of two energies; the average energy gain of the
electron in the laser moving over a laser wavelength, divided by its rest energy. The presence of the velocity
of light, $c$, signals the relativistic nature of the strength
parameter, with $a_0 > 1$ describing the regime where electrons become relativistic.

Note that $a_0$ is a purely
classical quantity as no factors of $\hbar$ are present. This should be compared with the situation when the denominator of (\ref{A0RATIO}) is written in terms of Sauter's critical field \cite{sauter:1931}, more commonly known as the Schwinger limit \cite{schwinger:1951}. This is defined by $m_e c^2=eE_\mathrm{crit}\lambdabar_C$, where $\lambdabar_C \equiv \hbar/ m_e c$ is the Compton wavelength.
At the critical field strength it becomes
energetically favourable for the vacuum to `break down' by
emitting electron positron pairs which will move apart such as
to decrease the external field until it becomes subcritical
again. This is clearly a quantum effect. Numerically, one finds $E_\mathrm{crit} = 1.3 \times 10^{18}$ V/m, which is
still five orders of magnitude above what can presently be
achieved. Inserting this definition into (\ref{A0RATIO}) we find $a_0 = (\lambdabar_L/\lambdabar_C)
(E_{\mathrm{rms}}/E_{\mathrm{crit}})$. The smallness of
the second factor is compensated by the large ratio of laser to
Compton wavelength which is of the order of $10^6$. That is why
the high-intensity regime, which is the main focus of this
paper, indeed corresponds to $a_0 > 1$, the current upper limit
being about $10^2$.

Looking at (\ref{A0DEF}) we note that $a_0$
appears to be a frame dependent quantity as the laser wavelength (and, more subtly, the time average involved in defining
$E_{\mathrm{rms}}$) do not represent Lorentz invariant
concepts. This shortcoming is claimed to be overcome
\cite{bamber:1999} by employing the gauge 4--potential $A_\mu$
to define $a_0$ via
\be \label{A0AA}
  a_0^2 = - \frac{e^2 \bra A_\mu A^\mu\ket}{m_e^2 c^4} \; .
\ee
This new definition, however, seems like a cure worse than the disease as (for a properly chosen average) it establishes Lorentz invariance by sacrificing gauge invariance -- (\ref{A0AA}) is not equal to (\ref{A0DEF}) in all gauges. This is clear from a particle physics perspective as $A_\mu A^\mu$ precisely corresponds to a
photon mass term (say in a Lagrangian) which certainly violates gauge invariance.

The main topic of this paper will hence be to provide a
universal definition of $a_0$ that meets both the requirements
of Lorentz and gauge invariance. Adopting common particle
physics practice we will set $\hbar = c = 1$ in the remainder
of this paper.

\section{\label{sec:2} Motion in plane wave fields}
Since we know from (\ref{A0DEF}) that $a_0$ is a classical quantity, it will suffice to consider the classical motion of a charged particle (chosen to be an electron throughout), as governed by the Lorentz equation
\be \label{LORENTZ}
  \dot{p}_\mu (\tau) = \frac{e}{m_e} \, F_{\mu\nu} \big( x(\tau)\big) \,  p^\nu (\tau) \;  ,
\ee
maintaining both gauge invariance and Lorentz covariance. Throughout, $p^\mu = m_e \dot{x}^\mu$ is the 4-momentum of the electron and dots represent derivatives with respect to proper time $\tau$.
The right-hand side of (\ref{LORENTZ}) is the Lorentz force
acting on the particle. Gauge invariance is
manifest as only the gauge invariant field strength tensor $F_{\mu\nu}$ appears\footnote{For a discussion of the gauge invariance of $a_0$ using $A_\mu$ see \cite{Reiss}.}.

Our first task is to solve (\ref{LORENTZ}) in terms of $F_{\mu\nu}$
rather than the potential $A_\mu$ (recall $F_{\mu\nu} =
\partial_\mu A_\nu - \partial_\nu A_\mu $) which was the
previously adopted approach \cite{kibble:1965a,berestetskii:1982} --  see, however, \cite{taub:1948}. For an arbitrary field configuration it will
generally not be possible to find an analytic solution of
(\ref{LORENTZ}) which is a \textit{nonlinear} differential
equation depending arbitrarily on $x(\tau)$ via $F_{\mu\nu}$. On the other hand, drastic simplifications arise if $F_{\mu\nu} (x)$ is a plane wave. In this case the field strength tensor may be written \cite{schwinger:1951}
\be  \label{FPWLIN}
F^{\mu\nu} (x) = F_1(k\cdot x) \, f_1^{\mu\nu} +F_2(k\cdot x)\,
f_2^{\mu\nu}\; ,
\ee
with the amplitudes $F_j$ depending on the Lorentz
invariant phase $k \cdot x$, where $k^\mu \equiv \omega n^\mu$ is
the 4-momentum of the wave. The constant tensors $f_j^{\mu\nu}$ are parameterised
in terms of dimensionless 4--vectors $n$ and $\epsilon_j$,
\be\label{FPWLIN2}
f_j^{\mu\nu}
\equiv n^\mu \epsilon_j^\nu - n^\nu \epsilon_j^\mu \; .
\ee
The lightlike vector $n_\mu$ may be chosen `orthogonal' to the spacelike polarisation
vectors $\epsilon_j$, i.e.\ $n \cdot \epsilon_j = 0$. For circular polarisation
we have $\epsilon_1\cdot \epsilon_2=0$ and for linear polarisation we simply set
one of the $F_j$ to zero. In either case we clearly have $n_\mu F^{\mu\nu} = 0$,  which expresses the transversality of the wave.
Hence, dotting $n$ (or $k$) into the equation of motion we find that the momentum  component $\dprod{k}{p}$ is a constant
of motion which, upon integrating, induces a proportionality between proper time
$\tau$ and `light--cone time' $n \cdot x$ (assuming $n\cdot x(0) = 0$).
For the phase $k\cdot x$ in (\ref{FPWLIN}) this implies $k \cdot x \equiv \Omega \tau$. Thus, quite remarkably, we can trade the $x$ dependence of $F^{\mu\nu}$ for a dependence solely on proper time $\tau$
whereupon the equation of motion becomes \textit{linear}. Consequently it
can be solved analytically by exponentiation. As the time dependence resides in
the scalar prefactors $F_j$, the tensor $F_{\mu\nu}$ commutes with itself at
different times and time--ordering of the exponential is unnecessary. Denoting the integral
of $e F_{\mu\nu}/m_e$ by $G_{\mu\nu}$ such that $G_{\mu\nu}(0) = 0$, the first integral of the equation of motion
(\ref{LORENTZ}) is
\be \label{PEXP}
	p_\mu (\tau) = \exp {\left[ G(\tau) \right]}_{\mu\nu} \, p^\nu(0) \; .
\ee
To evaluate the exponential in (\ref{PEXP}) we have
to determine all powers of $F_{\mu\nu}$ (or $G_{\mu\nu}$). For $F^2_{\mu\nu}$
one finds the rather simple result,
\be \label{FF}
	F^2_{\mu\nu} \equiv {F_\mu}^\sigma F_{\sigma\nu}= F_j F_j\, n_\mu n_\nu\; .
\ee
Multiplying with $F_{\mu\nu}$ once more,
transversality implies that its cube vanishes, $F^3_{\mu\nu} = 0$. Fields with this property are called null fields in general
relativity \cite{stephani:2004}. In Taub's classification of  constant fields
this case is denoted `parabolic' \cite{taub:1948} i.e.\ electric and magnetic
fields are orthogonal and of equal magnitude such that the basic bilinear
Lorentz and gauge invariants vanish,
\begin{align}
- \sfrac{1}{4} F_{\mu\nu} F^{\mu\nu} &= \sfrac{1}{2} (E^2 - B^2)
= 0 \; , \label{SCS} \\
- \sfrac{1}{4} F_{\mu\nu} \widetilde{F}^{\mu\nu} &= \vc{E} \cdot \vc{B} = 0 \; . \label{SCP}
\end{align}
Constant fields of this type are also
referred to as crossed fields, e.g.\ in \cite{berestetskii:1982}, and constitute
the simplest model of a laser beam applicable when all relevant length scales
are small compared to $\lambdabar_L$. As (\ref{SCS}) and (\ref{SCP}) are invariant statements, identifying fields as null is a Lorentz
invariant characterisation. It is easy to check that these identities are equally valid for the plane waves (\ref{FPWLIN}). We note in this context that the energy momentum tensor for null fields takes on a particularly simple form and coincides with (\ref{FF}) as the trace term vanishes, $ T_{\mu\nu} = F^2_{\mu\nu}$.

As a result of this discussion the exponential series
(\ref{PEXP}) is truncated to second order, giving the simple expression

\be \label{PPARA1}
  p_\mu (\tau) = \left[ g_{\mu\nu} +
  G_{\mu\nu}(\tau) + \sfrac{1}{2} G^2_{\mu\nu} (\tau) \right]
  p^\nu(0) \; ,
\ee
with $g_{\mu\nu}=\text{diag}(1,-1,-1,-1)$. This constitutes the gauge invariant first integral of the
Lorentz equation (\ref{LORENTZ}). It is a trivial matter to
integrate again, which yields a gauge invariant expression for the particle orbit. For the moment, (\ref{PPARA1}) is sufficient for a proper determination of the strength parameter
$a_0$.

From (\ref{A0AA}), where $a_0^2$ is quadratic in $A_\mu$, we expect a gauge invariant definition to be quadratic in $F_{\mu\nu}$ or
$G_{\mu\nu}$. Furthermore it should be related to an energy
density or intensity, which suggests using the energy momentum tensor $T_{\mu\nu}$. As the usual Lorentz invariants (\ref{SCS}) and (\ref{SCP}) vanish we are lead to the only remaining invariant $p_\mu T^{\mu\nu} p_\nu$ given by contracting with the gauge invariant momentum (\ref{PPARA1}). There is more than one way to make
this dimensionless, but our previous formulae (\ref{A0DEF}) and
(\ref{A0RATIO}) uniquely lead to the definition
\be \label{A0GEN}
  a_0^2 \equiv \frac{e^2}{m_e^2} \frac{\bra\!\bra p_\mu T^{\mu\nu} p_\nu \ket\!\ket}{(k
  \cdot p)^2} \; .
\ee
Here $\langle\!\langle\ldots\rangle\!\rangle$ denotes the Lorentz invariant proper time average which, in this paper, we take over {\it all} proper time such that it yields the Fourier zero mode.

The invariant numerator in (\ref{A0GEN}), originally introduced in \cite{NR} as $(F_{\mu\nu}p^\nu)^2$, is indeed proportional to the energy density seen by the charge, as in the charge rest frame, where $p = (m_e, \vc{0})$, one has $p_\mu
T^{\mu\nu} p_\nu = m_e^2 T^{00} = m_e^2 (E^2 + B^2)/2$. The general
relativistic analogue of this is an important characteristic of
curved space times, constrained by positive energy theorems
\cite{wald:1984}. The numerator of (\ref{A0GEN}) is then the average energy density seen by the charge during its \emph{entire} exposure to the laser field. The second invariant $\dprod{k}{p} = m_e \Omega$ in the denominator is the constant of motion proportional to the laser frequency seen by the particle. Altogether, $a_0$ is a ratio of two energies as described below (\ref{A0RATIO}).

Taking $p$ as the momentum of a probe photon
$a_0$ also determines the amount of vacuum birefringence in
ultra-strong laser fields due to an induced effective metric,
$h_{\mu\nu} = g_{\mu\nu} - \kappa \, T_{\mu\nu}$ with $\kappa =
\kappa(a_0)$ \cite{heinzl:2006}.

We conclude this section with three explicit examples: linearly and circularly polarised plane waves, and crossed fields.  In each case our definition (\ref{A0GEN}) recovers (\ref{A0DEF}), and the gauge non--invariant expression (\ref{A0AA}) is obtained from
(\ref{A0GEN}) by substituting for $F^{\mu\nu}$ in terms of the gauge
potential $A_\mu$ in light--cone gauge, $n \cdot A =0$ \cite{brown:1964}. We will now describe the orbits and calculate $a_0$ for our three examples. In all cases we choose initial conditions $\dprod{\epsilon_j}{p}(0)=0$, i.e. the electron has zero initial transverse momentum, and as before $\dprod{n}{x}(0)=0$. From (\ref{PPARA1}), we find that the orbits may be parameterised by three functions $b_1,b_2,b_3$. These are derived from the $F_j$ describing the laser fields as in (\ref{FPWLIN}) and (\ref{FPWLIN2}), and will be given below for each example. The form of the orbits is:
\be
\begin{split}
  x^\mu(\tau) =& x^\mu(0) + \frac{\tau}{m_e}p^\mu(0)
  - \frac{e\,b_j(\tau)}{\dprod{k}{p}}\, \epsilon_j^\mu +
  \frac{e^2\,b_3(\tau)}{2(\dprod{k}{p})^2}\, k^\mu \; .
\end{split}
\ee
The first two terms represent initial conditions and the final two terms describe the transverse and longitudinal motion, respectively. We proceed to the examples.\\
\textit{Linear polarisation}. We have $F_1 \equiv \omega C \cos
\dprod{k}{x}$, with $C$ a constant amplitude, and $F_2\equiv 0$. The orbit is described by $b_1=C(1-\cos \Omega\tau)$, $b_2=0$ and $b_3=C^2(2\Omega\tau-\sin{ 2\Omega\tau})/4$. With transverse coordinates $x$, $y$ and longitudinal coordinate $z$, we see a frequency doubling in the longitudinal orbit coefficient $b_3$. Accordingly, and in the frame where the average drift velocity vanishes, the motion in the $x$--$z$ plane acquires the well
known `figure-8' shape \cite{berestetskii:1982,mcdonald:1986}
which is nothing but the Lissajous figure corresponding to
frequency ratio 2:1, see Figure \ref{figs}. The energy-momentum
tensor is $T^{\mu\nu} =
C^2k^\mu k^\nu \cos^2 \Omega\tau$ with the cosine
averaging to $1/2$ so that $a_0^2 = e^2C^2/2 m_e^2$.

\textit{Circular polarisation}. We choose $F_1$ as above and  $F_2 \equiv -\omega C \sin \Omega\tau$. In this case $b_1$ is as above, $b_2=-C(\Omega\tau-\sin\Omega\tau)$ and $b_3 = -2C b_2$ i.e.\ there is no frequency doubling.
In the average rest frame where all drift velocities vanish the trajectory is an ellipse in the $x$--$z$ plane. The energy momentum tensor becomes constant, $T^{\mu\nu} = C^2 k^\mu k^\nu$, implying a laser amplitude $a_0 = eC/m_e$.

In both these examples our $a_0^2$ defines the effective mass of an electron in the field \cite{Sengupta} via the square of the average 4--momentum, $\langle\!\langle p\rangle\!\rangle^2=m_e^2\,(1+a_0^2)$, identifying ${1+a_0^2}$ with a gamma factor squared \cite{mcdonald:1986}. Periodicity of the above fields
results in periodic orbits, hence no net acceleration. This is
different for our final example.

\textit{Crossed fields}. These are obtained by letting $\Omega\to 0$ in either
of the cases above. We find $F_1=C\omega$ and $F_2=0$ implying constant field strength. Transverse and
longitudinal orbit coefficients become quadratic and cubic in $\tau$
respectively: $b_1=C\Omega^2\tau^2/2$, $b_2=0$, $b_3=C^2\Omega^3\tau^3/3$. As the charge does not
experience a full period of the laser field there is no cancellation between positive and negative field
regions and we hence have net acceleration (subcycle acceleration). The energy momentum tensor and $a_0$ are the same as for
circular polarisation.

\section{Electromagnetic vortices}

A more sophisticated null field model of a laser beam is the electromagnetic vortex found by Bia{\l}ynicki-Birula \cite{bialynicki-birula:2004}. The author states that this
approximates a circularly polarised Laguerre--Gauss beams in the
vicinity of the beam axis/vortex line. The tensor structure is as in (\ref{FPWLIN}) and (\ref{FPWLIN2}) but the amplitude functions now depend on transverse coordinates as well as on the phase $k\cdot x$.  It is useful to introduce complex notation for the amplitudes, $\mathscr{F} \equiv F_1 + iF_2$, the polarisation vectors, $\epsilon \equiv \epsilon_1 + i \epsilon_2$ and the transverse coordinates, $Z
\equiv \epsilon \cdot x$.

The vortex field strength tensor
\be
  F^{\mu\nu} = \sfrac{1}{2}\, \mathscr{F}
  (n^\mu \epsilon^\nu - n^\nu \epsilon^\mu) + c.c.\;,
\ee
is determined by the complex `envelope' function,
$
  \mathscr{F} \equiv C\omega\, Z \exp (-i \dprod{k}{x}) \; .
$
It follows that the energy momentum tensor is of type (\ref{FF}),
\be
  T^{\mu\nu} = |\mathscr{F}|^2 n^\mu n^\nu = C^2 \omega^2 |Z|^2n^\mu n^\nu\; ,
\ee
but now depends explicitly on the transverse coordinate $Z
(\tau)$. Nevertheless, this form implies that, according to
(\ref{A0GEN}), the dimensionless intensity parameter for
electromagnetic vortices is
\be
  a_0^2 = \frac{e^2 C^2}{m_e^2} \bra\!\bra\, ZZ^*\ket\!\ket \equiv
  \omega_c^2 \bra\!\bra\, ZZ^*\ket\!\ket \; ,
\ee
with cyclotron frequency $\omega_c$. To determine this elegant expression explicitly we obviously need to find $Z$ by solving the equation of motion,
\be
  \ddot{x}^\mu = \sfrac{1}{2} \omega \omega_c \,\dprod{\epsilon}{x}\left\{
  \dprod{\epsilon}{\dot{x}} \; n^\mu -
  \dprod{n}{\dot{x}} \; \epsilon^\mu
  \right\} + c.c. \; .
\ee	
\begin{figure}[!h]
\centering\includegraphics[width=2.6cm]{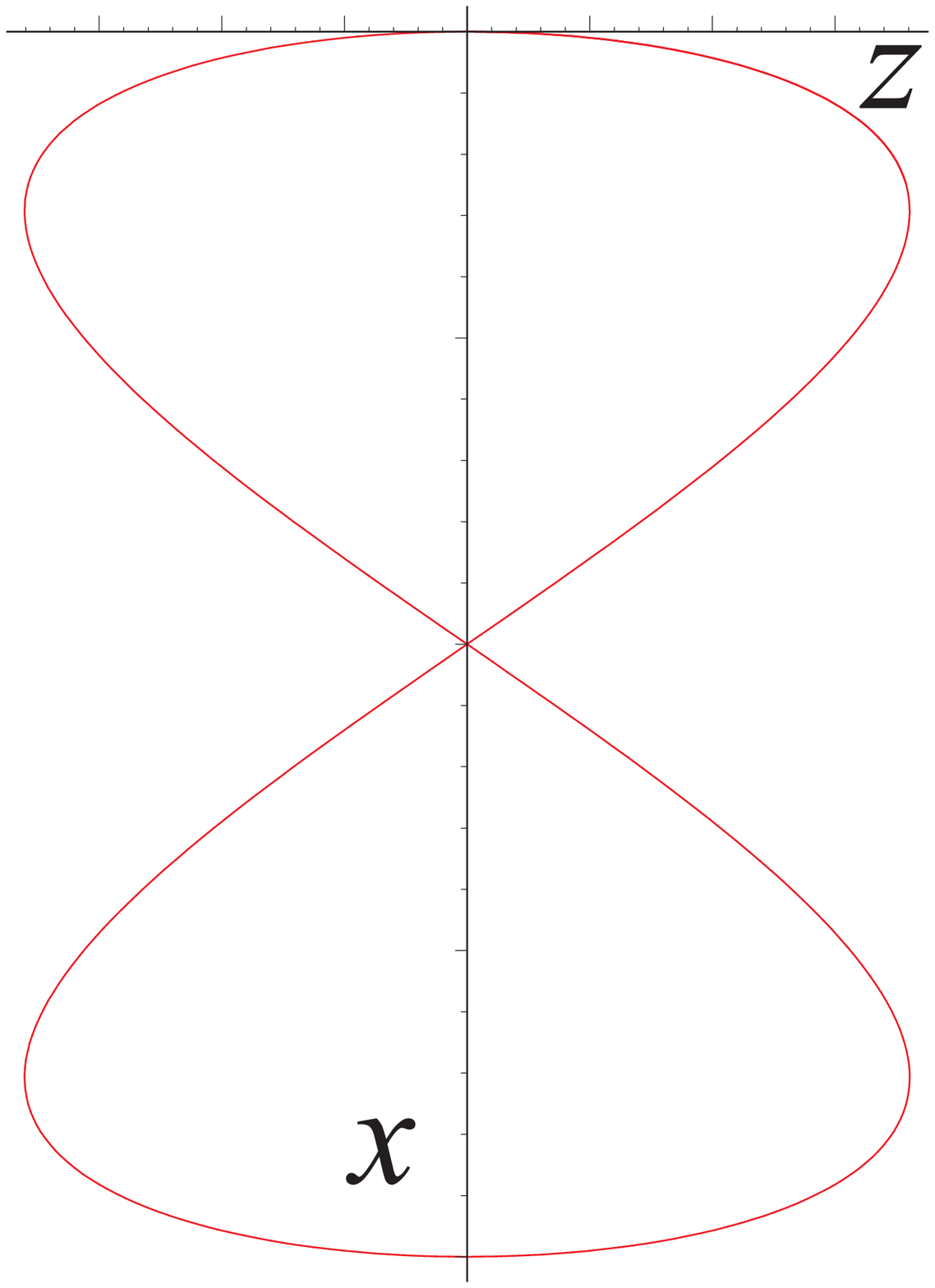}\hspace{20pt}\includegraphics[width=3.6cm]{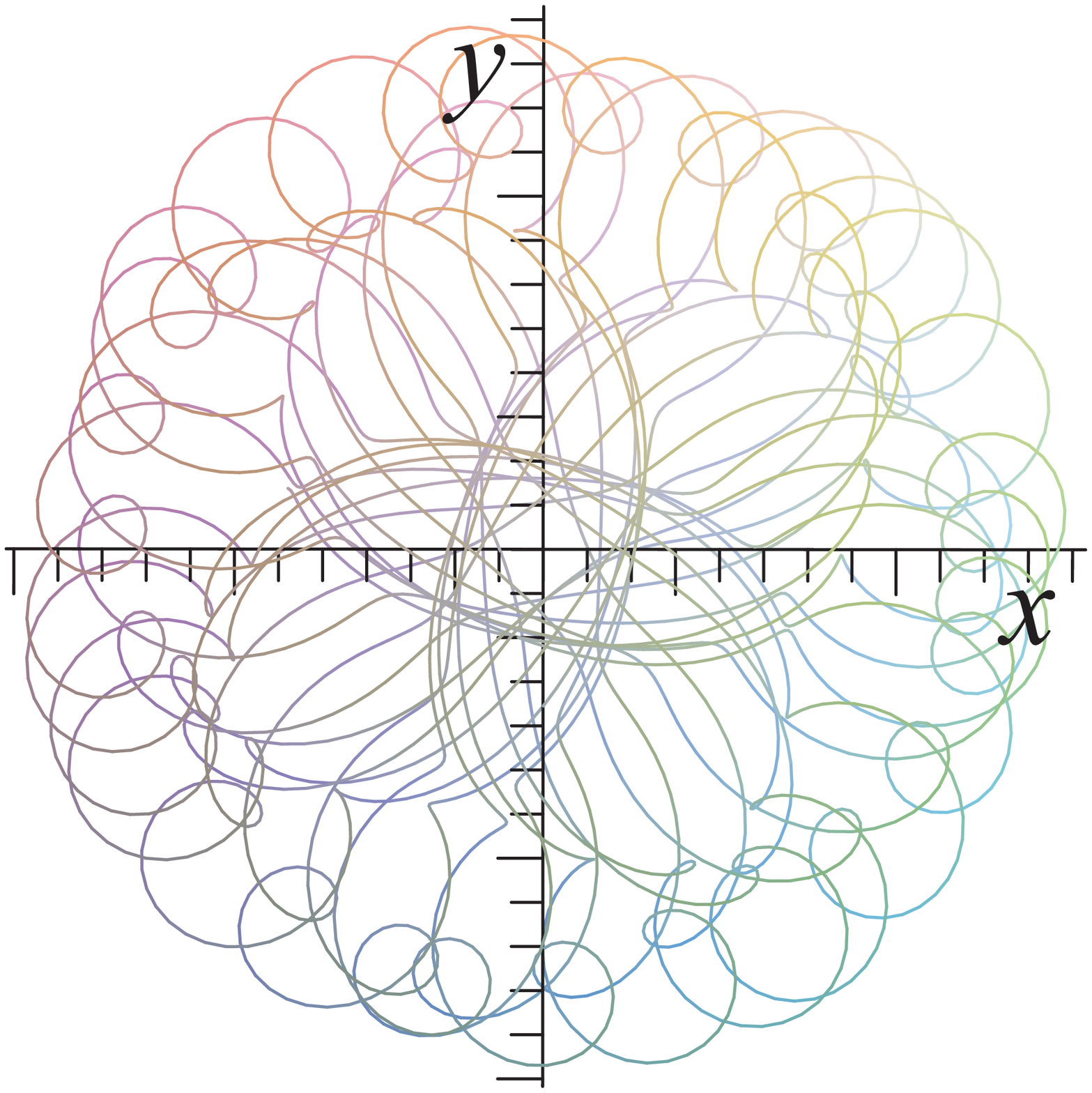}
\caption{Left: motion in a linearly polarised plane wave (electron's average rest frame). Right: bounded motion in the vortex, projected into the transverse plane.}
\label{figs}
\end{figure}
We point out that this equation is
\textit{nonlinear} and can\textit{not} be solved by exponentiation. Fortunately, Bia{\l}ynicki--Birula has shown there is still an exact
solution. He notices that longitudinal ($n^\mu$) and
transverse ($\epsilon^\mu$) motion decouple. The former is easily solved because the tensor structure again guarantees transversality so $k \cdot p$ is conserved exactly as before.  This conservation renders the transverse equation linear. Its solution amounts, in our notation, to introducing a co-moving basis $\epsilon_\tau$ via
\be \label{Z}
  \zeta (\tau) \equiv e^{-i\Omega\tau/2} \, Z(\tau)  \equiv \epsilon_\tau
  \cdot x \; .
\ee
In terms of $\zeta$ the transverse equation of motion is
\be
  \ddot{\zeta} + i \Omega \dot{\zeta} - \frac{\Omega^2}{4} \zeta - \omega_c \Omega\, \zeta^* = 0 \; ,
\ee
which is clearly linear, describing a damped harmonic
oscillator. Its solution is given explicitly in \cite{bialynicki-birula:2004}, so we do not reproduce it here. For $\kappa\equiv 4\omega_c/\Omega < 1$ it describes non--periodic bounded motion, see Figure \ref{figs}. $\kappa>1$ yields acceleration which, however, is an artifact of the unphysical transverse growth of the fields.

With the solution in hand, we return to $a_0$. It is clear from (\ref{Z}) that the intensity parameter becomes
\be
a_0^2 = \omega_c^2 \bra\!\bra \zeta\zeta^* \ket\!\ket\;,
\ee
and it remains only to evaluate the proper time average over transverse coordinates. We find the resulting $a_0$ to depend explicitly on the charge's motion through its initial position and momentum. e.g. if the particle is at rest initially at transverse co--ordinates $x_0$, $y_0$, the intensity parameter is
\begin{equation}
  a_0^2 = \frac{\omega_c^2}{2} \frac{x_0^2 \, (1+\kappa) \, (2-\kappa)
  + y_0^2 \, (1- \kappa) \, (2+ \kappa)}{(1- \kappa^2)} \;.
\end{equation}
This dependence on the particle's orbit is not so surprising: the same is
actually true for the plane wave, as $a_0$ contains $e$, $m_e$ and an average over the particle's proper time. As the vortex has a non--trivial spacetime dependence, any measure of its strength should reflect this.

\section{Discussion and conclusion}

We have given a thorough discussion of particle motion in null
fields from a particle physics perspective. The fields in
question are regarded as simple models of laser fields, for which the usual Lorentz and gauge
invariants quadratic in electromagnetic fields vanish. It
follows that the laser fields can only be characterised in
terms of a particle \textit{probe}. Its 4--vector $p$ then may
be employed to form a nontrivial invariant $p_\mu T^{\mu\nu}
p_\nu$ which is the energy density of the field as seen by the
probe in its instantaneous rest frame. By making this quantity
dimensionless one arrives at a universal definition
of laser strength satisfying the crucial requirements
of both Lorentz and gauge invariance. For more complicated
beams $a_0$ becomes dependent on
transverse coordinates and hence acquires a profile. Its
nonvanishing gradient will in turn determine the ponderomotive
force.

\acknowledgments
The authors thank G.~Dunne, C.~Harvey, K.~Langfeld, M.~Lavelle, D.~McMullan, V.~Serbo and A.~Wipf for useful discussions.


\begin{thebibliography}{99}

\bibitem{strickland:1985}
A.~Strickland and G.~Mourou, Opt.\ Commun. {\bf 56}, 212 (1985).

\bibitem{Geddes:2004tb}
S.~P.~D.~Mangles {\it et al.}, Nature {\bf 431} (2004) 535;
  C.~G.~R.~Geddes {\it et al.}, {\it ibid.} 538;
  J.~Faure {\it et al.}, {\it ibid.} 541.

\bibitem{woodward:1946}
P.~Woodward, J.~IEE {\bf 93}, 1554 (1946), Part IIIA;
P.~Woodward and J.~Lawson, J.~IEE {\bf 95}, 363 (1948), Part III.

\bibitem{palmer:1987}
R.~Palmer, (1987), SLAC-PUB-4320.

\bibitem{bamber:1999}
C.~Bamber {\em et~al.}, Phys.~Rev.~D {\bf 60}, 092004 (1999).

\bibitem{Dunne:2004nc}
  G.~V.~Dunne,
  [arXiv:hep-th/0406216].

\bibitem{sauter:1931}
F.~Sauter,
\newblock Z.\ Phys. {\bf 69}, 742 (1931).

\bibitem{schwinger:1951}
J.~Schwinger,
\newblock Phys.~Rev. {\bf 82}, 664 (1951).

\bibitem{Reiss}
H.~R.~Reiss, Phys.~Rev.~A \textbf{19}, 1140 (1979).

\bibitem{kibble:1965a}
T.~Kibble, Phys.~Rev. {\bf 138}, B 740 (1965);
E.~S. Sarachik and G.~T. Schappert, Phys. Rev. {\bf D1}, 2738 (1970);
A.~L. Troha {\em et~al.}, Phys. Rev. {\bf E60}, 926 (1999).

\bibitem{berestetskii:1982}
V.~Berestetskii et al,
\newblock {\em Quantum Electrodynamics (Course of Theoretical Physics)} (Butterworth-Heinemann, 1982).

\bibitem{taub:1948}
A.~Taub,
\newblock Phys.~Rev. {\bf 73}, 786 (1948).

\bibitem{stephani:2004}
H.~Stephani,
\newblock {\em Relativity} (Cambridge Univ. Press, 2004).


\bibitem{NR}
A.~I.~Nikishov and V.~I.~Ritus, Sov.\ Phys.\ JETP {\bf 19}, 529 (1964).

\bibitem{wald:1984}
R.~Wald,
\newblock {\em General Relativity} (Univ. Chicago Press,
  1984).

\bibitem{heinzl:2006}
T.~Heinzl {\em et~al.},
\newblock Opt.~Commun. {\bf 267}, 318 (2006).

\bibitem{brown:1964}
L.~Brown and T.~Kibble,
\newblock Phys.~Rev. {\bf 133}, A705 (1964).


\bibitem{mcdonald:1986}
K.~T. McDonald, (1986), preprint DOE/ER/3072-38;
www.hep.princeton.edu/\~{}mcdonald/e144/prop.pdf.


\bibitem{Sengupta}
N.~D.~Sengupta, Bull.~Math.~Soc. (Calcutta) \textbf{44}, 175 (1952).


\bibitem{bialynicki-birula:2004}
I.~Bia{\l}ynicki-Birula,
Phys.\ Rev.\ Lett. {\bf 93}, 020402 (2004).



\end{thebibliography}
\end{document}